\documentclass[
    ,final            
    ,tnotealph
  ]
  {aipproc}

\layoutstyle{6x9}

\begin{document}

\title{Accretion and diffusion in the DAZ white dwarf GALEX~J1931+0117}


\author{St\'ephane Vennes}{
  address={Astronomick\'y \'ustav, Akademie v\v{e}d \v{C}esk\'e republiky, 
  Fri\v{c}ova 298, CZ-251 65 Ond\v{r}ejov, Czech Republic}
}

\author{Ad\'ela Kawka}{
  address={Astronomick\'y \'ustav, Akademie v\v{e}d \v{C}esk\'e republiky, 
  Fri\v{c}ova 298, CZ-251 65 Ond\v{r}ejov, Czech Republic}
}

\author{P\'eter N\'emeth}{
  address={Astronomick\'y \'ustav, Akademie v\v{e}d \v{C}esk\'e republiky, 
  Fri\v{c}ova 298, CZ-251 65 Ond\v{r}ejov, Czech Republic}
}

\begin{abstract}
We present an analysis of high-dispersion and
high signal-to-noise ratio spectra of the DAZ white dwarf
GALEX~J1931+0117. The spectra obtained with the VLT-Kueyen/UV-Visual Echelle Spectrograph
show several well-resolved Si\,{\sc ii} spectral lines enabling a
study of pressure effects on line profiles. We observed large
Stark shifts in silicon lines in agreement with laboratory
measurements. A model atmosphere analysis shows that
the magnesium, silicon and iron abundances
exceed solar abundances, while the oxygen and calcium abundances are below solar.
Also, we compared the observed line profiles to synthetic spectra computed with variable accretion
rates and vertical abundance distributions assuming diffusion steady-state.
The inferred accretion rates vary from $\dot{M}=2\times10^6$ for calcium to $2\times10^{9}$ g\,s$^{-1}$ for
oxygen and indicate that the accretion flow is dominated by oxygen, silicon and iron while being deficient in carbon, magnesium and calcium.
The lack of radial velocity variations between two
measurement epochs suggests that GALEX~J1931+0117 is probably not in a close binary and that
the source of the accreted material resides in a debris disc.
\end{abstract}

\maketitle


\section{Introduction}

GALEX~J193156.8+011745 (GALEX~J1931+0117, thereafter) is a hydrogen-rich white dwarf
\citep{ven2010} showing optical heavy-element lines and an infrared excess.
The original low-resolution spectrum obtained with the
New Technology Telescope (NTT) at La Silla Observatory showed a strong Mg\,{\sc ii}$\lambda$4481 doublet and weaker
silicon lines. Follow-up echelle spectroscopy obtained with the Very Large Telescope (VLT)-Kueyen
enabled a detailed abundance study. The near-solar ($\pm0.5$ dex) abundances of oxygen, magnesium, silicon, calcium
and iron bear the signature of an external supply of material accreting onto the surface of the white dwarf.
Based on available data, Vennes et al. concluded that the supply may originate from a close,
sub-stellar companion or from a cool debris disc.

The presence of heavy elements in hydrogen-rich white dwarfs has variously been interpreted as
intrinsic to the star, or as extrinsic, i.e., supplied by the interstellar medium \citep{dup1993}, by a nearby companion as in
post-common envelope systems \citep{deb2006,kaw2008}, or by a debris disc \citep{zuc2003,kil2006,far2008}.
However, accretion from the  interstellar medium is unlikely because of supply shortages \citep{far2010a}.
In the extrinsic scenarios, the elements are accreted and diffused in the atmosphere and envelope of the star
\citep[see][]{fon1979,koe2009}.
An intrinsic, or internal, reservoir of heavy elements is also possible, but in either scenario
a self-consistent solution of the diffusion equation must explore the effect
of radiative acceleration on trace elements \citep{cha1995a,cha1995b}.

As a class, the polluted DA white dwarfs, or DAZs, are often defined by the
detection of the Ca\,{\sc ii} H\&K doublet in optical spectra \citep{zuc2003,koe2005}.
Exceptionally, Mg\,{\sc ii}$\lambda$4481 is, so far, only detected in a handful of warm white dwarfs \citep[but see][]{kaw2010}
such as EG~102 \citep{hol1997}, GALEX~J1931+0117, and two warm white dwarfs from the Sloan Digital Sky Survey (SDSS)
that show evidence of dusty and gaseous discs \citep{gan2006,gan2007}. The presence of a
large concentration of magnesium in the last two objects helped establish a strong link between heavy element
pollution and dense circumstellar environments.
Moreover, an infrared excess that cannot otherwise be explained by a cool companion, may
be attributed to a dust ring as in the case of the DAZ white dwarf G29-38 \citep{gra1990}.

We present new high-dispersion spectroscopic observations that help elucidate the nature of the peculiar abundance
pattern in GALEX~J1931+0117. We revise our abundance measurements and explore the effect of
vertical abundance inhomogeneities. 

\section{Optical spectroscopy and abundances}

\begin{figure}
\includegraphics[width=0.70\textwidth]{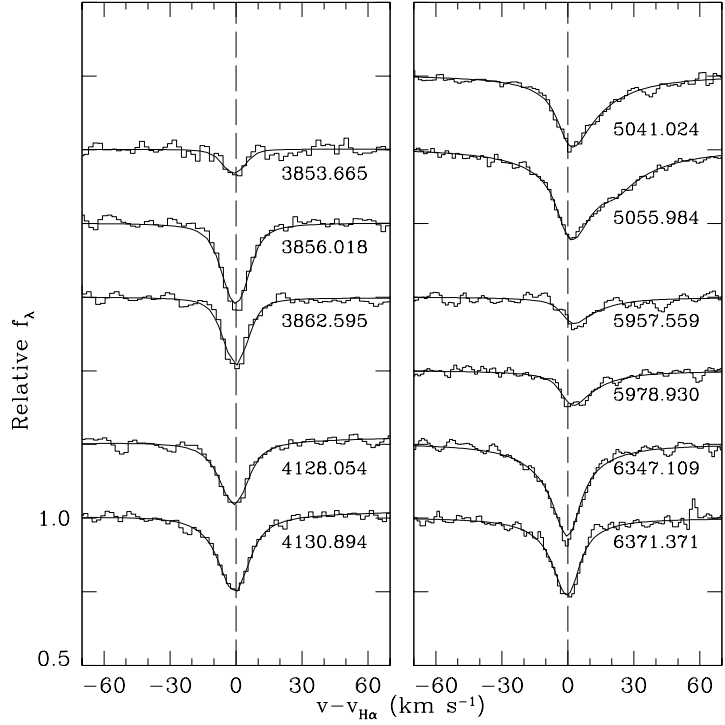}
\caption{Set of eleven Si\,{\sc ii} lines obtained with the VLT/UVES. The line profile analysis
reveals an abundance of silicon nearly a factor of two above solar. The Si\,{\sc ii}$\lambda\lambda$5041,5055,
5957,5978 lines
show pressure (Stark) shifts in agreement with laboratory measurements.
}
\label{fig1}
\end{figure}

Following the identification of GALEX~J1931+0117 as a DAZ white dwarf we obtained a series of
echelle spectra ($R\approx 46000$) using standard settings with the UV-Visual Echelle Spectrograph (UVES) attached to the VLT-Kueyen.
The first series of spectra were obtained on UT 2009 Nov 12 (epoch 1, thereafter) with the dichroic \#2
and the ``HER\_5'' and ``BK7\_5'' filters on the blue and red arms, respectively.
The blue spectrum was centred at 4370\AA\ and covered the range 3757-4985\AA, and the
red spectra were centred at 7600\AA\ and covered the ranges 5698-7532 and 7660-9464\AA.
The first series of spectra were analysed by us \citep{ven2010}.
The second series of spectra were obtained on UT 2010 Mar 15 (epoch 2, thereafter) with the dichroic \#1
and the ``HER\_5'' and ``SHP700'' filters on the blue and red arms, respectively.
The blue spectrum was centred at 3900\AA\ and covered the range 3290-4518\AA, and
the red spectra were centred at 5800\AA\ and covered the ranges 4788-5750 and
5839-6808\AA.

\begin{figure}
\includegraphics[width=0.70\textwidth]{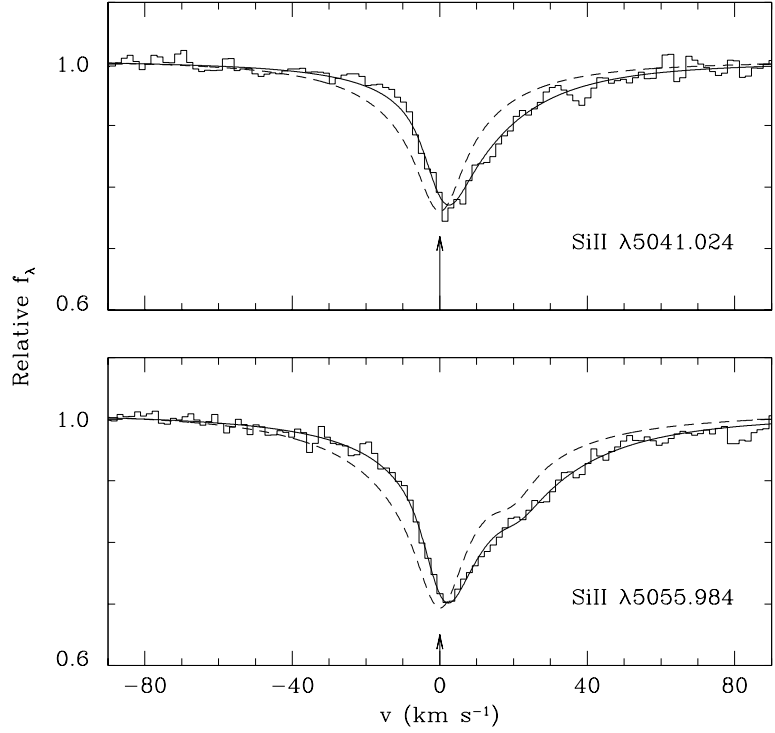}
\caption{Details of Si{\sc ii} line profiles showing an asymmetry and a velocity shift. Model including
pressure (Stark) shifts (full lines) are markedly superior to models excluding it (dashed lines).
}
\label{fig2}
\end{figure}

The H$\alpha$ radial velocity measured at both epochs are identical 
within errors. We adopted the average $\textsl{v}_{\rm H\alpha}=37.2\pm0.6$\,km\,s$^{-1}$.
The lack of radial velocity variations rules-out the 
presence of a close companion and supports the debris disc model.
Figure~\ref{fig1} shows Si\,{\sc ii} lines with the velocity scale centred on $\textsl{v}_{\rm H\alpha}$.
Stark shifts \citep{gon2002,les2009} are apparent in some lines. 
Figure~\ref{fig2} compares observed line spectra to the model line profiles that include or exclude Stark shifts.
The Stark shifts result in a line asymmetry and a notable velocity shift.
Details of the calculations are given by \citet{ven2011}.
Table~\ref{tbl1} summarizes our new abundance measurements based on a critical review of atomic data
\citep{ven2011}.

\section{Accretion and diffusion in the atmosphere}

In the accretion/diffusion model the abundance at each layer is obtained by solving simultaneously
the mass-continuity and diffusion equations.
The mass-continuity equation for element ``2'' is
\begin{displaymath}
\dot{M_2} = 4\pi\,R^2\,(A_2\,n_2) \textsl{v}_{1,2}
\end{displaymath}
where $R$ is the stellar radius, and
$\dot{M_2}$, $A_2$, $n_2$, and $\textsl{v}_{1,2}$ are the mass accretion rate, the atomic weight, the number density, 
and the diffusion velocity of element ``2'', respectively. The diffusion velocity is
given by the diffusion equation. Assuming diffusion steady-state the mass-continuity equation
is solved for the abundance ratio
\begin{displaymath}
c_2 \equiv \frac{n_2}{n_1} \approx \dot{M_2} \frac {1}{4\pi R^2}\frac{1}{A_2}\frac{\tau_{1,2}}{m},
\end{displaymath}
where $c_2<<1$ is the abundance of trace element ``2'' relative to the main constituent ``1''
assumed to be hydrogen
and $\tau_{1,2}$ is the diffusion time scale of element ``2'' related to the diffusion velocity by
\begin{displaymath}
\tau_{1,2} = \frac{m}{\rho\, \textsl{v}_{1,2}},
\end{displaymath}
where $m$ and $\rho$ are the total mass loading and density in the atmosphere. The selective
radiation  pressure and concentration gradients are negligible in the present case and the diffusion velocity is given by
\begin{displaymath}
\textsl{v}_{1,2} \approx D_{1,2} (1-A_2) \frac{m_p\,g}{kT},
\end{displaymath}
where $D_{1,2}$ is the diffusion coefficient of element ``2'' which is a function of temperature, density and atomic charge.
The diffusion coefficient for ionized species is given by \citet{fon1979} and includes a correction factor for
thermal diffusion.
\citet{mon1976} proposed to calculate the diffusion coefficient for neutral species based on the hard-sphere model:
\begin{displaymath}
D_{1,2}({\rm h-s}) = 1.36\times10^{19} \sqrt{\frac{T(A_2+1)}{A_2}} \frac{1}{\sigma_0^2 n_p}  \ {\rm cm}^2\,{\rm s}^{-1}
\end{displaymath}
where $T$ is the temperature in K, $n_p$ is the proton density in cm$^{-3}$, and 
$\sigma_0$ is the atomic radius (in \AA) taken as 0.91, 0.65, 1.72, 1.46, 2.23, and 1.72 for C, O, Mg, Si, Ca, and Fe, respectively.
However, \citet{mic1978} showed that the neutral-proton interaction is dominated by the induced electric dipole and
that the hard-sphere model may overestimate the diffusion coefficient by a large factor.  
The polarization diffusion coefficient is approximated by
\begin{displaymath}
D_{1,2}({\rm pol}) = 3.3\times10^{16}\,T\,\sqrt{\frac{A_2+1}{A_2 \alpha_D}}\frac{1}{n_p} \ {\rm cm}^2\,{\rm s}^{-1}
\end{displaymath}
where $\alpha_D$ is the dipole polarizability (in \AA$^3$). We adopted averaged values of 1.68, 0.86, 10.6, 5.49, 23.6, and 8.95 for
C, O, Mg, Si, Ca, and Fe, respectively from the tabulation of P. Schwerdtfeger (http://ctcp.massey.ac.nz/dipole-polarizabilities).
For example, at a reference temperature of $20\times10^3$ K, the ratio of hard-sphere to polarization diffusion coefficients is
$D_{1,2}({\rm h-s})/D_{1,2}({\rm pol})\approx 3$ for Mg, Si, Ca, and Fe, $\approx 5$ for C, and $\approx 6$ for O.
The diffusion coefficient averaged over ionization species is given by:
\begin{displaymath}
D_{1,2} = \sum_{i} x_i\, D_{1,2,i}
\end{displaymath}
where $x_i=n_{2,i}/n_2$ are the ionization fractions for element ``2'' and $D_{1,2,i}$ are the corresponding diffusion
coefficients.

\begin{figure}
\includegraphics[width=0.70\textwidth]{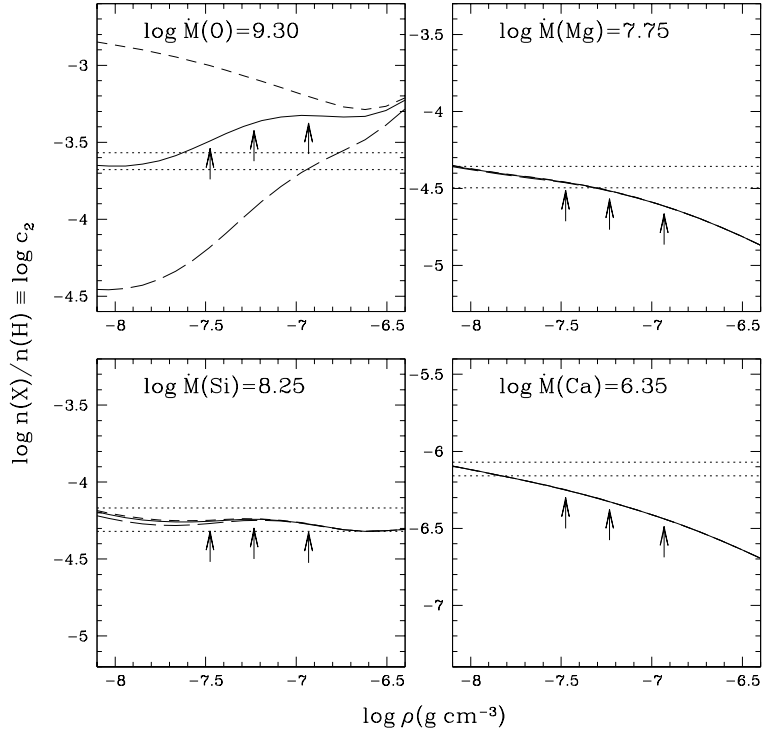}
\caption{Abundance profiles in the line forming region predicted by the accretion-diffusion model. The accretion
rates that fit the observed O, Mg, Si, and Ca line spectra are given in g\,s$^{-1}$. The abundances measured using homogeneous
model atmospheres are shown with dotted horizontal lines. Diffusion profiles excluding neutral species (short dashed lines) overestimate
the abundances in the line forming region relative to models including neutral species and realistic polarization diffusion coefficients (full lines), while models using
hard-sphere diffusion coefficients (long dashed lines) underestimate the abundances.
Mass densities corresponding to optical depth unity at $\lambda=7400$ (leftmost), 4700, and 1260 \AA\ (rightmost) are marked with vertical
arrows.
}
\label{fig3}
\end{figure}

Figure~\ref{fig3} shows the predicted abundance profiles of O, Mg, Si, and Ca in the atmosphere of GALEX~J1931+0117 calculated
using the procedure described above. The accretion rates were set at the best-fitting values. The effect of neutral species on the
profiles of Mg, Si, and Ca is negligible because of their relatively high ionization fractions. On the other hand, the predicted oxygen
abundance profile in the
line forming region is dominated by diffusion of neutral oxygen atoms. Employing the hard-sphere 
diffusion coefficient for
neutral oxygen leads to an overestimation of the diffusion velocity and an underestimation of the
abundance by up to a factor of 4 relative to more realistic calculations that
employ the polarization diffusion coefficient as determined by \citet{mic1978}.

\begin{table}
\centering
\begin{minipage}{85mm}
\caption{Abundances and accretion rates in GALEX~J1931+0117.}
\label{tbl1}
\centering
\begin{tabular}{ccccc}
\hline
X & $n({\rm X})/n({\rm H})$  & [X/H] \tablenote{[X/H]$\equiv\log{n({\rm X})/n({\rm H})}-\log{n({\rm X})/n({\rm H})}_\odot$} & $\log\,\dot{M}$ (g\,s$^{-1}$)  &  [X/Si] \tablenote{[X/Si]$\equiv\log{n({\rm X})/n({\rm Si})}|_{\rm flow}-\log{n({\rm X})/n({\rm Si})}_\odot$} \\
\hline
C     & $<7\times10^{-5}$         & $<-0.54$ & $<8.0$         & $-0.8$ \\
O     & $2.4\pm0.3\times10^{-4}$  &  $-0.28$ & $9.30\pm0.05$ & $+0.1$ \\
Mg    & $3.8\pm0.6\times10^{-5}$  &  $+0.05$ & $7.75\pm0.06$  & $-0.5$ \\
Si    & $5.8\pm1.0\times10^{-5}$  &  $+0.25$ & $8.25\pm0.07$  & ...    \\
Ca    & $7.7\pm0.8\times10^{-7}$  &  $-0.42$ & $6.35\pm0.04$  & $-0.9$ \\
Fe    & $3.7\pm0.8\times10^{-5}$  &  $+0.12$ & $8.50\pm0.09$  & $+0.0$ \\
\hline
\end{tabular} 
\end{minipage}
\end{table}

Table~\ref{tbl1} lists the accretion rates $\dot{M}$ for each element measured by fitting spectral syntheses based on
calculated abundance profiles to the observed line spectra. The inferred abundances in the accretion flow
reveal a relatively high abundance of oxygen, silicon, and iron relative to carbon, magnesium and calcium.

\section{Spectral energy distribution}

\begin{figure}
\includegraphics[width=0.65\textwidth]{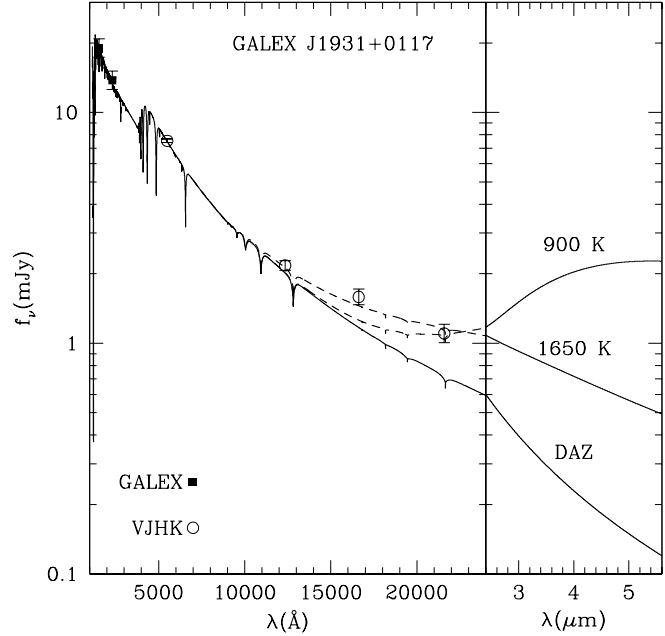}
\caption{Predicted spectral energy distribution of GALEX~J1931+0117 (full lines) extending beyond the JHK
photometry and including the emergent flux of a hot white dwarf (DAZ) added to black-body emission from
a disc at 900 or 1650 K. The white dwarf dominates the ultraviolet region ({\it GALEX} $N_{\rm UV}$ and
$F_{\rm UV}$) while the disc is predicted to dominate the infrared region.
}
\label{fig4}
\end{figure}

A lack of radial velocity variations suggests that the source of the accreted material is a
circumstellar debris disc.
Simulated infrared spectral energy distributions for GALEX~J1931+0117 including
a spectral synthesis of the white dwarf added to a warm disc (Fig.~\ref{fig4}) show that the
observed distribution \citep{ven2010} can be matched by a
warm disc (900 K) or a hot disc (1650 K) near sublimation temperature.
The inferred dust
temperature must be below sublimation temperatures \citep[see][]{lod2003} for the main constituents such as calcium (1659 K), or
silicon and magnesium (1354 K).
The corresponding disc sizes are 90$\Omega_{WD}$ for the hot disc, and, compensating for
its reduced emissivity per unit area, 2200$\Omega_{WD}$ for the warm
disc, where $\Omega_{WD}$ is the solid angle subtended by the white dwarf itself.
The observed spectral energy distribution extends to the 2MASS measurements in the infrared
and measurements at longer wavelengths are required to further constrain the model.

A comparison with other warm DAZ white dwarfs supports these conjectures. The DAZ PG1015+161
\citep{koe2006} is similar
to GALEX~J1931+0117, and, based
on JHK and Spitzer observations, \citet{jur2007} inferred the presence of a
disc with an inner temperature ranging from 800 to 1000 K and emitting area
from 700 to 1200 $\Omega_{WD}$. In a similar case, \citet{far2009} inferred the presence of a disc with a blackbody temperature of 1500K around PG1457-086. 
The great diversity in disc emissivity is illustrated by the case of
HS0047+1903. Despite having a similar calcium abundance, no evidence of a disc is
found in JHK and Spitzer observations \citep{far2010b}.

\section{Conclusions}

We presented a model atmosphere analysis of the high-metallicity white dwarf GALEX~J1931+0117.
The abundance pattern obtained using homogeneous model atmospheres shows that magnesium, silicon, and iron
are near or above solar abundances,
while carbon, oxygen and calcium are below solar abundances. However,
a line profile analysis performed using vertical abundance distributions obtained by solving the
steady-state diffusion equation shows that the accretion flow is rich in oxygen, silicon, and iron while
it is depleted in other elements. Although the oxygen abundance is below solar in the
line forming region, it must be supplied in larger quantity because of its short
diffusion time-scale relative to other elements.

The effect of pressure shifts are apparent in several strong silicon lines. This effect 
predicted by \citet{ham1989} and \citet{krs1992} also impacts radial velocity measurements.
A lack of radial velocity variations between two epochs that are 123 days apart
rules out a close binary scenario
for the origin of the accreted material. We are left with the possibility that the infrared excess
belongs to a warm disc of debris material that accretes onto the white dwarf surface.

In conclusion, our analysis of the abundance pattern in GALEX~J1931+0117
and the absence of a close companion, as well as a comparison with similar cases (e.g., DAZ PG1015+161)
support the likely presence of a dusty disc around GALEX~J1931+0117.

\begin{theacknowledgments}
This research is supported by GA AV grant numbers IAA301630901 and
IAA300030908, respectively, and by GA \v{C}R grant number P209/10/0967.
\end{theacknowledgments}

\bibliographystyle{aipproc}   

\end{document}